\documentclass[published]{JHEP3}
\JHEP{02(2004)023}

\usepackage{epsfig}

\skip\footins = 1\bigskipamount plus 2pt minus 4pt

\newcommand{\SU}{\mathop{\rm SU}\nolimits}

\def\gsi{\gtrsim}
\newcommand{\gsim}{\mathop{\gsi}}
\newcommand{\nn}{\nonumber}
\newcommand{\eps}{\epsilon}
\newcommand{\la}{\langle}
\newcommand{\ra}{\rangle}
\newcommand{\vnu}{\vert \nu \vert}

\title{Axial correlation functions in the $\eps$-regime: a numerical
  study with overlap fermions}

\author{Wolfgang Bietenholz and \ Stanislav Shcheredin\\
  Institut f\"{u}r Physik, Humboldt Universit\"{a}t zu Berlin \\
  Newtonstr.\ 15, D-12489 Berlin, Germany \\
  E-mail: \email{bietenho@physik.hu-berlin.de},
          \email{shchered@physik.hu-berlin.de}}

\author{Thomas Chiarappa, Karl Jansen and Kei-ichi Nagai\\
  NIC/DESY Zeuthen \\ 
  Platanenallee 6, D-15738 Zeuthen, Germany\\
  E-mail: \email{chiarapp@ifh.de},
          \email{karl.jansen@desy.de},
          \email{knagai@ifh.de}}

\abstract{We present simulation results employing overlap fermions for
  the axial correlation functions in the $\epsilon$-regime of chiral
  perturbation theory. In this regime, finite size effects and
  topology play a dominant r\^{o}le.  Their description by quenched
  chiral perturbation theory is compared to our numerical results in
  quenched QCD.  We show that lattices with a linear extent $L > 1.1
  \,{\rm fm}$ are necessary to interpret the numerical data obtained in
  distinct topological sectors in terms of the $\epsilon$-expansion.  Such
  lattices are, however, still substantially smaller than the ones
  needed in standard chiral perturbation theory.  However, we also
  observe severe difficulties at very low values of the quark mass, in
  particular in the topologically trivial sector.}

\received{November 25, 2003}
\accepted{February 9, 2004}
\keywords{Lattice QCD, Chiral Lagrangians}

\begin{document}

\section{Introduction}\label{section1}

There are a number of questions in QCD where lattice simulations are
hardly feasible in the physical regime, e.g., when the light quark
masses assume their physical values and the physical volume is large
enough to accommodate the light mesons. However, in some occasions it
is possible to extract physical information from simulation results
obtained in an \emph{unphysical} setup, in particular in an
unphysically small volume.  The variation of the volume then allows
for the investigation of finite size effects (FSE), which provide
information about the same system in a large volume.  Examples for
this strategy are the non-perturbative renormalization procedure based
on the Schr\"{o}dinger Functional~\cite{SF} and the evaluation of a
scattering length based on the analysis of the level
repulsion~\cite{scat}. A further example --- which will be discussed
here --- is the study of the dynamics of quasi-Goldstone bosons in the
$\eps$-regime of chiral perturbation theory ($\chi$PT).

One procedure to make use of the FSE is the transmutation of the
variation of the volume into a variation of the physical scale.
Then the effect on certain scale-dependent observables enables
for instance the determination of a running coupling constant~\cite{runcoup}.

An alternative procedure --- that we are going to deal with --- is
based on the fact that even in small volumes the FSE are characterized
by \emph{infinite volume parameters}. Therefore the evaluation of FSE
provides access to physical quantities.  That situation occurs in
particular in a specific regime of $\chi$PT, where the pion Compton
wave length is much larger than the linear extent of the volume, while
the inverse linear extent is far below the QCD cutoff scale.  The
resulting FSE are described in $\chi$PT --- which considers the pions
as quasi-Goldstone bosons of the chiral symmetry breaking --- by the
so-called $\eps$-expansion~\cite{GasLeu}.  The corresponding
zero-modes are treated exactly by means of collective variables, while
the non-zero modes, as well as the pion mass, are dealt with
perturbatively; note that the fluctuations around the zero modes do
fit into the box. The domain where this expansion applies is denoted
as the \emph{$\eps$-regime}.

As a test case for QCD, the $\eps$-expansion can also be applied to
the $O(N)$ symmetric linear~\cite{Neulin} and non-linear~\cite{HasLeu}
$\sigma$-model.  For the case of the 4d $O(4)$ model the numerical
technique of extracting infinite volume parameters from FSE was
applied very successfully~\cite{O4}. In QCD such theoretical concepts
are often tested in the quenched approximation where simulations are
much easier, although the concept is designed for full QCD. It turned
out that quenched results are quite often close to experimental
quantities~\cite{quench}, even though there is no reason to expect
this behavior in general. At low energy, it is important that also
\emph{quenched $\chi$PT} has been worked
out~\cite{qCPT1}--\cite{PHDnew}, hence quenched QCD can be used to
probe the applicability of this procedure.  This is the motivation for
our pilot study, which has the purpose to gain first experience on
\begin{itemize}
\item the range for the physical volume where the $\eps$-expansion is
  applicable
\item the extent of the statistics which is required for reliable
  results in this framework
\item the sources of potential difficulties, and their dependence on
  the topological sector.
\end{itemize}
Our statistics is rather modest and it should be enlarged for precise
results. We hope, however, to provide a sound basis for the prospects
of simulations in the $\eps$-regime, also in view of full QCD.

So far the $\eps$-regime appeared as a \emph{terra incognita} for
numerical studies of QCD because until recently no formulation of
lattice fermions was known which would be suitable in this
context. The obstacles for such a formulation are linked to the
problems in formulating chiral symmetry and topological charges $\nu$
on the lattice --- note that topology becomes extremely important for
$\chi$PT in small volumes~\cite{LeuSmi}.  None of these problems is
solved by the Wilson fermion, and at small quark masses the latter
also runs into technical problems such as the significance of
exceptional configurations. The staggered fermion, on the other hand,
seems to be ``topology blind'', i.e.\ its results do not depend on
$\nu$ --- at least not for strong and moderate gauge couplings, which
were tested so far --- in contrast to the
continuum~\cite{stagger}.\footnote{Using very smooth gauge fields and
  smeared links, topological structures can be seen by the staggered
  fermion, as a very recent study shows in the Schwinger
  model~\cite{DH}.}

A solution to these two problems is now available due to the recent
formulation of chirally invariant lattice fermions. Their Dirac
operator obeys the Ginsparg-Wilson relation~\cite{GW} and an operator
of that kind will be used in this work. However, its application is
very computation time demanding, hence its use in QCD is restricted to
the quenched approximation at present and in the foreseeable future.
For domain wall fermions as another realization of a Dirac operator
inducing an exact lattice chiral symmetry, an attempt had been made to
explore the $\epsilon$-regime~\cite{PO}. However, the lattices used in
that work were unfortunately too small for $\chi$PT to work.

In section~\ref{section2} we describe the lattice formulation of our
simulations and the formulae of quenched $\chi$PT which are relevant
in this context.  In section~\ref{section3} our numerical results for
the axial correlation functions in distinct topological \pagebreak[3] sectors are
compared to the predictions of quenched $\chi$PT.  We also discuss the
stability of such numerical results, which is most problematic in the
topologically trivial sector.  Section~\ref{section4} is devoted to
our conclusions and to an outlook on this new field of research.  A
synopsis of our results was given in refs.~\cite{KIN,TC}.

\section{Chiral invariant lattice fermions and chiral perturbation theory}\label{section2}

\subsection{Overlap fermions}\label{section2.1}

The lattice Dirac operator that we are going to use is the so-called
overlap operator~\cite{Neu}.  On a lattice of unit spacing the
massless operator is given by 
\begin{equation} 
\label{over} 
D_{\rm ov}^{(0)} = \mu  \left[ 1 +
{A}/{\sqrt{A^{\dagger}A}}\right], \qquad A = D_{W} - \mu \,,
\end{equation} 
where $D_{W}$ is the Wilson-Dirac operator (with Wilson parameter $1$
and $\kappa = 1/8$), and $\mu$ is a mass parameter, which can be
chosen in some interval that shrinks as the gauge coupling
increases. Our simulations were performed at $\beta =6$ and we set
$\mu = 1.4$, which is optimal for locality in this
case~\cite{HJL}. This observation referred to the Wilson gauge action
that we were also using throughout this work.

$D_{\rm ov}^{(0)}$ is a solution to the Ginsparg-Wilson
relation~\cite{GW}
\begin{equation}  \label{GWR}
D_{\rm ov}^{(0)} \gamma_{5} + \gamma_{5} D_{\rm ov}^{(0)} =
\frac{1}{\mu} D_{\rm ov}^{(0)} \gamma_{5} D_{\rm ov}^{(0)} \,,
\end{equation}
which excludes additive mass renormalization and exceptional
configurations.  Hence $D_{\rm ov}^{(0)}$ has exact zero modes with a
definite chirality~\cite{Has}. This provides a sound definition of the
topological charge of a lattice gauge configuration by means of the
Atiyah-Singer Index Theorem.  These amazing properties are related to
the fact that Ginsparg-Wilson fermions have an exact lattice chiral
symmetry~\cite{ML}, which turns into the standard chiral symmetry in
the continuum limit.

For practical applications we still have to insert a small bare quark
mass $m_{q}$. It is added to $D_{\rm ov}^{(0)}$ as follows,
\begin{equation}  \label{over_m}
D_{\rm ov} = \Big( 1 - \frac{m_{q}}{2 \mu} \Big) D_{\rm ov}^{(0)} +
m_{q} \,.
\label{Dovm}
\end{equation}

The challenge to numerical studies is the inverse square root in
eq.~(\ref{over}). It is basically equivalent --- but somewhat more
convenient --- to evaluate the sign function of the hermitean operator
$\gamma_{5} A$. In our simulations we approximated this sign function
by Chebyshev polynomials to an accuracy of $10^{-12}$\, for the
eigenvalues and $10^{-16}$\, for the propagators.  For the description
of highly optimized algorithmic tools to deal with overlap fermions we
refer to refs.~\cite{Wupp,CERN}.

\subsection{The $\eps$-regime in quenched chiral perturbation theory}\label{section2.2}

The low energy properties of QCD can be described by an effective
lagrangian ${\cal L}_{\rm eff}$, and $\chi$PT is a powerful tool for
its construction\pagebreak[3] and evaluation~\cite{Leff}. At leading order, ${\cal
  L}_{\rm eff}$ is parametrized by the pion decay constant $F_{\pi}$
and the scalar condensate $\Sigma$,
\begin{equation}
{\cal L}_{\rm eff}[U] = \frac{F_{\pi}^{2}}{4} {\rm Tr} \left[
  \partial_{\mu} U \partial_{\mu} U^{\dagger} \right] - \frac{1}{2}
\Sigma m_{q} {\rm Tr} \left[ U e^{i\theta /N_{f}} + U^{\dagger}
  e^{-i\theta /N_{f}} \right],
\end{equation}
where $U(x) \in \SU(N_{f})$ is the quasi-Goldstone boson field,
$N_{f}$ is the number of quark flavors and $\theta$ the vacuum angle.
(For simplicity we assume a single bare mass $m_{q}$ for all flavors.)
Note that the values of $F_{\pi}$ and $\Sigma$ in a finite volume and
in infinite volume coincide, up to logarithmic corrections which are
specific to the quenched approximation.

Depending on the counting rules for the expansion, one distinguishes
two regimes for the $\chi$PT: the ``$p$-expansion'' applies to large
volumes where FSE are suppressed~\cite{p-exp}, which is the standard
situation.  In contrast, the ``$\eps$-expansion'' assumes a very small
quark mass --- even below the physical masses of the $u$ and $d$
quarks --- so that $m_{\pi} L \ll 1$, where $m_{\pi} \propto
\sqrt{m_{q}}$ is the pion mass and $L$ is the linear size of the the
system.  In this regime FSE are strong, but they can be described
analytically for dynamical as well as quenched
quarks~\cite{GasLeu,qCPT1,qCPT2}.  As an important feature, the
observables depend significantly on the topology~\cite{LeuSmi}, hence
they should be measured by averages over gauge configurations in a
fixed topological sector (or, more precisely, at a fixed absolute
value of the topological charge $\nu$). In the light of the
requirements of small quark masses and a well-defined topology, the
use of Ginsparg-Wilson fermions is ideal, as we pointed out before.

In this work we aim at a first interpretation of lattice data ---
obtained from simulating overlap fermions --- in the framework of the
$\eps$-expansion. In particular, our reference point is the formula
for the axial correlation function in quenched $\chi$PT derived in
ref.~\cite{qCPT2}.  Using this formula, the lattice data yield
predictions for the leading (quenched) low energy constants in the
effective lagrangian.  In order to stay in the $\eps$-regime we have
to require the dimensionless scaling variable
\begin{equation}  \label{zdef}
z = m_{q} \Sigma V
\end{equation}
to be small, i.e.\ clearly below $1$  (where $V$ is the volume).\footnote{If we insert
  $\Sigma = (250 \, {\rm MeV})^{3}$ as a typical value from the
  literature, the parameters of our results in figures~\ref{fig3}
  to~\ref{fig6} correspond to $z \simeq 0.34$.}

To render this paper self-contained we now quote the main result of
quenched $\chi$PT which we are going to use in our data analysis.
This result was obtained~\cite{qCPT2} consistently from two different
methods called ``supersymmetric''~\cite{SUSY} and
``replica''~\cite{replica}. Ref.\ ~\cite{qCPT2} also showed that the
corresponding vector correlation function vanishes to all orders in
the quenched $\eps$-expansion; a general argument is added in
Ref.~\cite{vec0}.

We start from the definition of the bare axial current at momentum
$\vec p = \vec 0$ and Euclidean time $t$,
\begin{equation} 
\label{axialcurrent}
A_{\mu}(t) = \sum_{\vec x} \bar \psi (t,\vec x ) \gamma_{5}
\gamma_{\mu} \psi (t, \vec x) \,.
\end{equation}
The formula for the axial correlation function in a volume $L^{3}
\times T$, to the first order in quenched $\chi$PT, reads\footnote{We
  have an extra factor of 2 compared to ref.~\cite{qCPT2} because our
  definition of the current does not involve the flavor space generators.}
\begin{eqnarray}  
\la A_{0}(t) \, A_{0}(0) \ra_{\nu} &=& 
2\cdot \left(\frac{F_{\pi}^{2}}{T} 
 + 2 m_{q} \Sigma_{\vert \nu \vert}(z) 
T  \cdot h_{1}(\tau )\right), 
\nonumber\\
h_{1}(\tau ) &=& \frac{1}{2} \left( \tau^{2} - \tau + \frac{1}{6} \right),
\qquad \tau = \frac{t}{T}\,, 
\nn \\
\Sigma_{\nu}(z) &=& \Sigma \left( z \left[ I_{\nu}(z) K_{\nu}(z) + 
I_{\nu +1}(z) K_{\nu -1}(z) \right] + \frac{\nu}{z} \right).
\label{AA}
\end{eqnarray}
$I_{\nu}$ and $K_{\nu}$ are modified Bessel functions, $z$ is defined
in eq.~(\ref{zdef}), $\nu$ is again the topological charge, and
$h_{1}$ is a purely kinematic function. The latter shows that this
correlation function is given by a parabola in $t$ with its minimum at
$T/2$. This behavior is qualitatively different from the infinite
volume, where the correlations decay exponentially (resp.\ as a
$\cosh$ function in a periodic volume).  However, the small and the
infinite volume have in common that in both cases the quenched axial
correlation function to the first order only depends on the coupling
constants $F_{\pi}$ and $\Sigma$.  In contrast, the quenched
pseudoscalar-pseudoscalar and the scalar-scalar correlation functions
involve further parameters~\cite{qCPT1}, which motivated our focus on
the simple case of the axial correlation function~(\ref{AA}).  We note
that the bare axial current in eq.~(\ref{axialcurrent}) needs to be
renormalized multiplicatively by applying a renormalization constant
$Z_A$, see below.

At this point we would like to mention that Random Matrix Theory can
be applied to QCD~\cite{OTV}, also in the quenched case~\cite{PHD}.
Random Matrix Theory involves additional assumptions and it leads to
explicit predictions for the distributions of the low lying
eigenvalues of the Dirac operator~\cite{RMT}. If these predictions
hold, they provide another way to extract values for $\Sigma $ and
(somewhat more involved) for $F_{\pi}$ from lattice data.  In recent
studies of the spectrum of $D_{\rm ov}^{(0)}$ it turned out that these
Random Matrix predictions are essentially confirmed, \emph{if} the
volume is not too small, which means for the linear size: $L \gsim
1.2~{\rm fm}$~\cite{BJS}.  This agrees roughly with an earlier study
using a truncated fixed point action on a $4^{4}$ lattice~\cite{Bern}.
Proceeding to larger physical volumes the quality of the agreement
with the Random Matrix prediction improves further~\cite{RMTfollowup}.

In particular the probability distribution for the first non-zero
Dirac eigenvalue $\vert \lambda_{1}\vert $ has a peak, which moves to
larger values (of $\vert \lambda_{1} \vert V \Sigma$) as $\vnu$
increases, where $\nu$ is the topological charge.  In the neutral
sector, $\nu = 0$, there is a significant density of very small
eigenvalues. A previous study of the chiral
condensate~\cite{condensate} observed that this property yields severe
problems for measurements in the sector $\nu =0$.  That observation
was confirmed in a striking way by our investigation that we are going
to present in the next section.

\section{Numerical results}\label{section3}

In our simulations we used the overlap operator $D_{\rm ov}$ described
in subsection~\ref{section2.1}.  We simulated quenched QCD on periodic
lattices of two sizes, $10^{3} \times 24$ and $12^{4}$, both at $\beta
=6$. This corresponds to physical volumes of $(0.93\,{\rm fm})^{3}
\times 1.86 \,{\rm fm}$ resp.\ $(1.12~{\rm fm})^{4}$.\footnote{Our
  physical scale is based on the Sommer parameter~\cite{Sommer}. In
  the following we do not keep track of possible errors in that
  scale.}

To determine the topological charge we computed the index, i.e.\ the
difference between the number of right-handed and left-handed zero
modes of $D_{\rm ov}^{(0)}$, by using either the Ritz functional
method~\cite{ritz} or the Arnoldi procedure~\cite{arpack}.  We mention
in passing that the index was always given by the number of zero modes
with only one chirality; we never encountered zero modes of opposite
chiralities in the same configuration, as it is expected in general
for interacting chiral fermions~\cite{CERN,privcom}.

Our statistics on the two lattice sizes is given in table~\ref{stati}.
On such small volumes most configurations have charges 
$\vert \nu \vert \leq 2$, and we have the largest statistics at
$\vert \nu \vert =1$. 

\TABLE[t]{\centerline{\begin{tabular}{|c|c|c|c|c|}
\hline
lattice & complete &
\multicolumn{3}{|c|}{number of configurations} \\
size & statistics & $\nu = 0$ & $\vert \nu \vert = 1$ & 
$\vert \nu \vert = 2 $\\
\hline
\hline
$10^{3} \times 24 $ & 65 & 20 & 24  & 17 \\
\hline 
$12^{4}$ & 157 & 47 & 78 & 24 \\
\hline
\end{tabular}}%
\caption{The statistics of our simulations on two lattices,
which we simulated both at $\beta = 6$.\label{stati}\label{tab1}}}

The parameter $F_{\pi}$ is determined through the minimum of the
function in eq.~(\ref{AA}) at $\tau = 1/2$ and it is hence easy to
fit.  $\Sigma$, on the other hand, is related to the curvature of the
predicted parabola and more difficult to extract, as we are going to
see.\footnote{This situation is somehow complementary to the
  comparison of the eigenvalue distributions to their Random Matrix
  predictions, where $\Sigma$ is easier to determine.}

\FIGURE[t]{\epsfig{file=small-lat.epsi,angle=270,width=.7\textwidth,clip=}%
\caption{Data for the axial correlation function from a $10^{3}\times
  24$ lattice at $\beta =6$.  For comparison we also show the curves
  predicted by quenched $\chi$PT for $\Sigma = 0$ (solid line) and
  $\Sigma = (250 \, {\rm MeV})^{3}$ (dashed line). The plot for $\vnu
  =1$ (on the right) illustrates clearly that the lattice size chosen
  here is \emph{not} compatible with $\chi$PT, as it could be expected
  also based on the comparison to Random Matrix Theory
  \cite{BJS}.\label{small-lat}\label{fig1}}}

\FIGURE[t]{\epsfig{file=axhist-0-and-1.epsi,angle=270,width=.7\textwidth,clip=}%
\caption{Histories of the axial correlation function 
$A_{0}(t) A_{0}(0)$ on a $10^{3} \times 24$ lattice at
$\beta =6$ and $m_{q} = 21.3 ~{\rm MeV}$. We show the histories 
at $t=8$ and at $t=16$.
For some $\nu =0$ configurations (on the left) we recognize
marked spikes. At $\vnu =1$ (on the right)
the history is significantly smoother.\label{history}\label{fig2}}}

\FIGURE[t]{\centerline{\epsfig{file=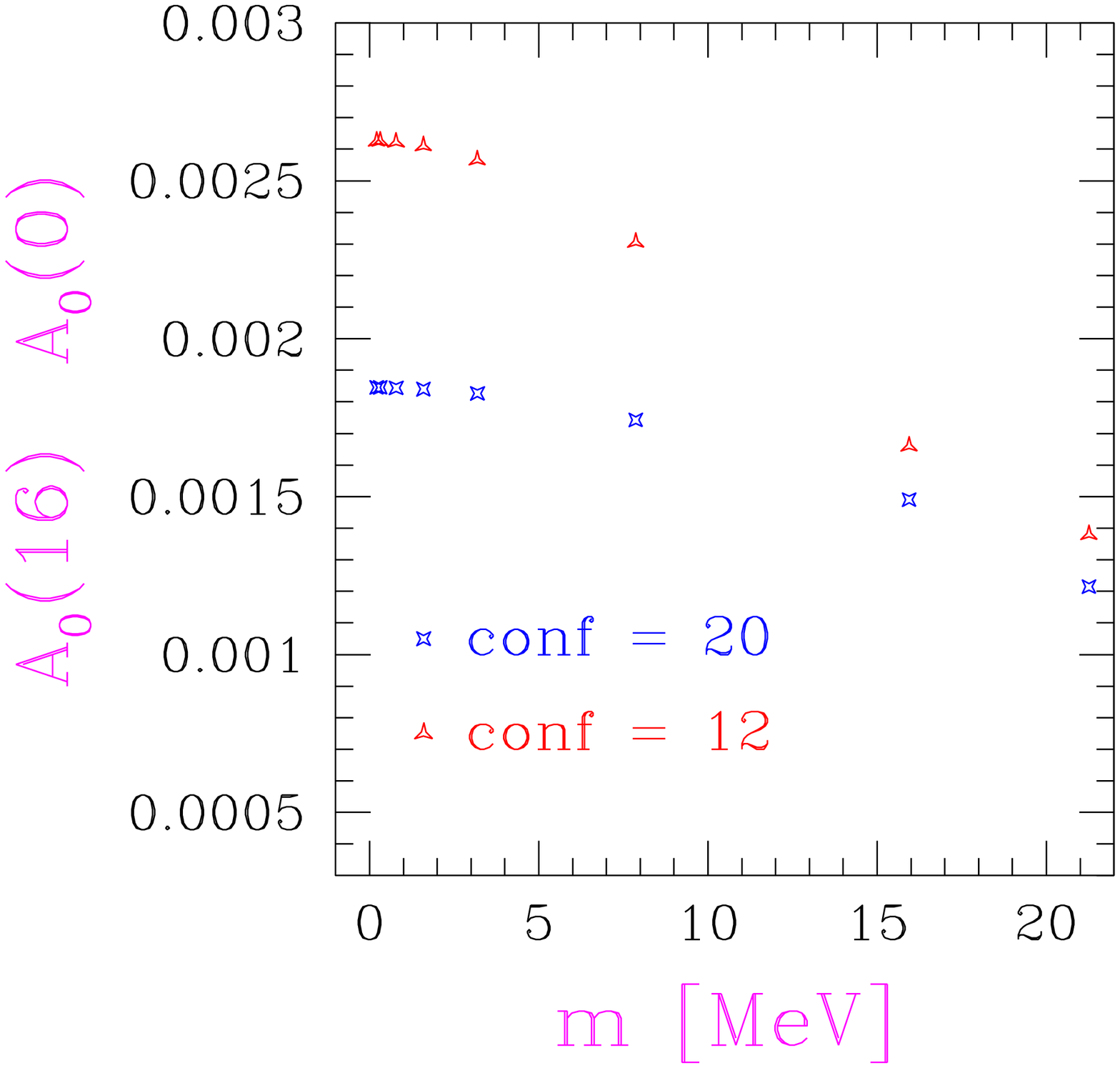,width=.4\textwidth,clip=}}%
\caption{The height of the two spikes of $A_{0}(16) A_{0}(0)$
in the $\nu =0$ history of figure~\ref{history}, at different quark 
masses $m_{q} = 0.2 \, {\rm MeV} \dots 21.3 \, {\rm MeV}$.\label{spikes}\label{fig3}}}

First we observed also here that the physical volume must exceed the
lower limit mentioned in subsection~\ref{section2.2}.  As a striking
example, we show that otherwise one does not obtain the curvature
required at $\vnu =1$ by any positive value of $\Sigma$.  This is
illustrated in figure~\ref{small-lat} (on the right), where our data
from the $10^{3}\times 24$ lattice are shown, as well as the analytic
curves given by eq.~(\ref{AA}) for $\Sigma = 0$ and $\Sigma = (250 ~
{\rm MeV})^{3}$.  The fact that those two curves can hardly be
distinguished is typical for all sectors with $\nu \neq 0$, due to the
last term in the expression of eq.~(\ref{AA}) for $\Sigma_{\nu}$.

Once we are in the $\epsilon$-regime, $\nu =0$ may look like the most
convenient case from eq.~(\ref{AA}).  In particular this is the only
sector which does not have the problem of the tiny sensitivity of the
predicted function on the value of $\Sigma$, see
figure~\ref{small-lat} on the left.  Unfortunately, from the numerical
point of view this sector is a nightmare. Its problems are related to
the danger of very small eigenvalues, as we mentioned in
subsection~\ref{section2.2}. This causes the large error bars in
figure~\ref{small-lat} on the left.  To illustrate this problem
explicitly we show in Fig.\ \ref{history} the histories for our
measurements of $\la A_{0}(t) \, A_{0}(0) \ra_{0}$ and $\la A_{0}(t)
\, A_{0}(0) \ra_{1}$.  In the topologically neutral sector there are
strong spikes.  We checked that these spikes appear exactly at those
configurations which have very small eigenvalues.  The height of the
spikes is maximal at very small bare quark mass $m_{q}$, as
figure~\ref{spikes} shows.\footnote{Figure~\ref{spikes} shows data for
  bare masses in the range of $m_q = 0.2\,{\rm MeV} \dots 21.3\,{\rm
    MeV}$.  The corresponding overlap operators $D_{\rm ov}$ could be
  inverted simultaneously at all these masses by means of a Multiple
  Mass Solver.  However, if one involves very small masses the latter
  has a problem of accuracy~\cite{Pilar}, hence we did not rely on it
  for our results at $m_{q} = 21.3\,{\rm MeV}$ and used standard
  inverters at this single mass only.\label{MMS}} In the chiral limit
$m_{q} \to 0$ the spike height is proportional to $1/\vert
\lambda_{1}\vert$. Qualitatively this property also holds for the
(smaller) spikes in the non-trivial sectors, where, however, the
occurrence of such very small modes is suppressed; this suppression
gets stronger for increasing values of $\vnu$.  Therefore we have some
kind of a tuning problem for $m_{q}$: on one hand it must be very
small to make sure we are in the $\eps$-regime, but on the other hand
taking it too small makes measurements statistically tedious, in
addition to numerical problems when evaluating the propagators (c.f.\
footnote~\ref{MMS}).

Returning to $m_q = 21.3 \, {\rm MeV}$, we would now like to get an
idea of the effect of such spikes, i.e.\ we want to estimate how many
configurations would be required to obtain stable expectation
values. For this purpose we consider the contribution of the smallest
(non-zero) eigenvalue $\lambda_{1}$ alone to the scalar
condensate. This contribution reads
\begin{equation}  \label{Smineq}
\Sigma_{\rm min}^{(\nu )} = \frac{1}{V} \int_{0}^{\infty} dz_{1} \,
P_{\nu}(z_{1}) \, \frac{2m_{q}}{m_{q}^{2} + (z_{1}/\Sigma V)^{2}} \ ,
\quad z_{1} = \lambda_{1} \Sigma V \,,
\end{equation}
where we insert the probabilities $P_{\nu}(z_{1})$ provided by Random
Matrix Theory~\cite{RMT}. We generated fake numbers for the lowest
eigenvalue with probability $P_{\nu}(z_{1})$ and computed $\Sigma_{\rm
  min}^{(\nu )}$ from them. The result can only be trusted when the
standard deviation becomes practically constant.  Figure~\ref{Smin}
shows the evolution of this standard deviation as the statistics is
enhanced. We infer that the sector $\nu =0$ requires a tremendous
statistics of $O(10^{4})$ configurations, but the situation is clearly
better for $\nu \neq 0$, where $O(100)$ configurations might be
sufficient.

\FIGURE[t]{\epsfig{file=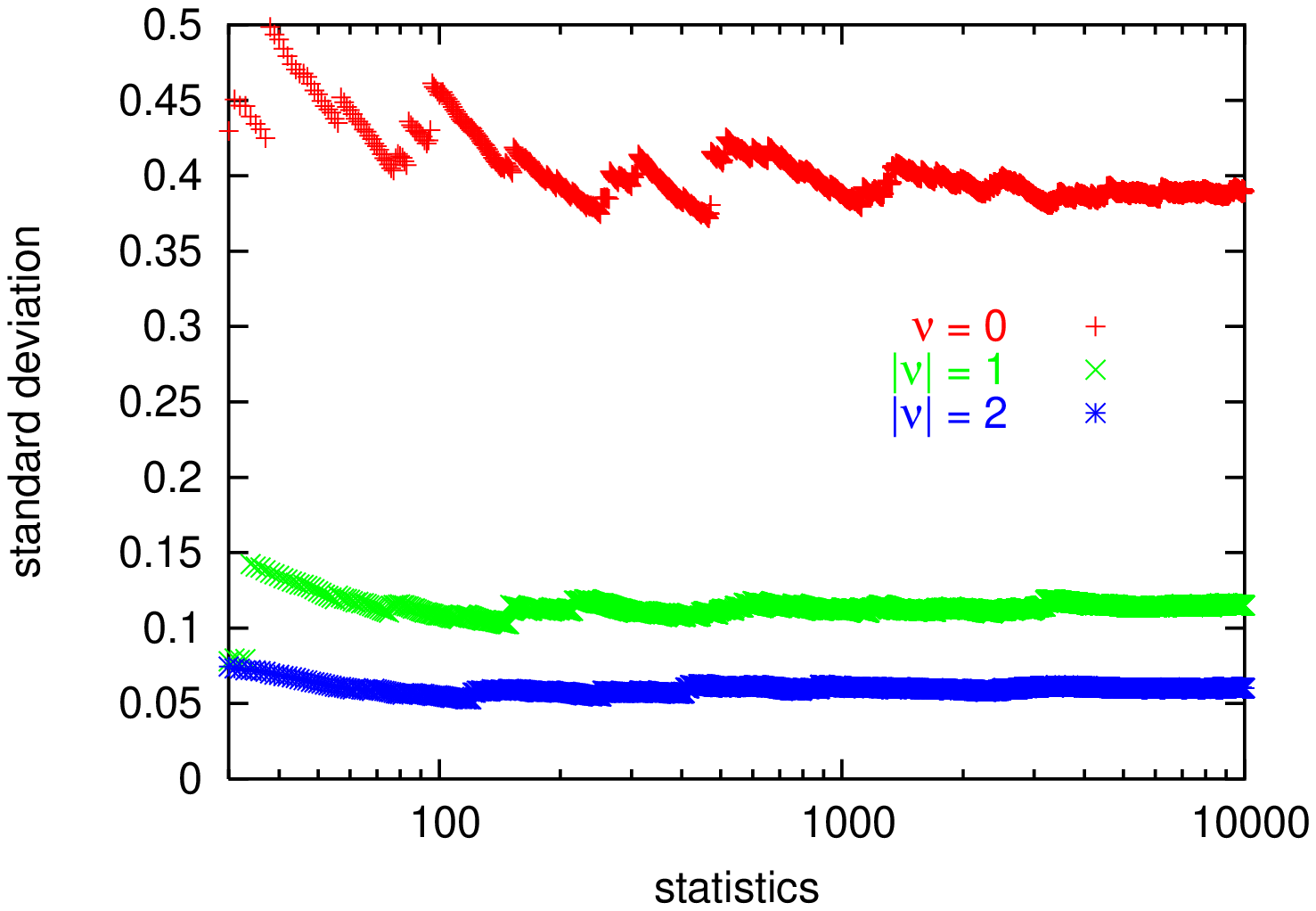,angle=270,width=.6\textwidth,clip=}%
\caption{The standard deviation in the (fake) measurement of
  $\Sigma_{\rm min}^{(\nu )}$, the contribution of the lowest non-zero
  eigenvalue to the scalar condensate, given in eq.~(\ref{Smineq}).
  The parameters correspond to a $12^{4}$ lattice at $\beta =6$ and
  $m_{q} = 21.3 \, {\rm MeV}$, and we assumed $\Sigma = (250\, {\rm
    MeV})^{3}$.\label{Smin}\label{fig4}}}

We now present our results from the $12^4$ lattice, which is close to
the point where the applicability of $\chi$PT should set in.  The
quark mass amounts again to $m_{q} = 21.3\,{\rm MeV}$.

Figure~\ref{axiFig} (on the left) shows our data for the axial
correlation function at $\vnu = 1$. The curve corresponding to eq.\
(\ref{AA}) can be fitted well over some interval that excludes the
points near the boundary (which are strongly affected by the
contributions of excited states).  This determines the additive
constant --- and therefore $F_{\pi}$ --- quite accurately. In
figure~\ref{axiFig} on the right we plot the resulting values of
$F_{\pi}$ as a function of the number $t_f$ of $t$ values (around
$T/2$) that we include in the fit. We find a decent plateau, which
suggests a value of $F_{\pi} = (86.7 \pm 4.0) \, {\rm MeV}$ for the
quenched, bare pion decay constant.  Although the purpose of the
present paper is mainly to demonstrate that measurements of $\la
A_0(t) A_0(0) \ra_{\nu \neq 0}$ do have the potential to determine
$F_{\pi}$, let us shortly address the question of renormalization. In
principle, the renormalized axial correlation function
$$
\la A_0^{R}(t) A_0^{R}(0) \ra
\equiv Z_{A}^2\cdot \la A_0(t) A_0(0) \ra
$$ 
has to be considered. As an estimate of the renormalization constant
$Z_{A}$ the value determined in ref.~\cite{ZA} can be taken which
amounts to $Z_{A}\simeq 1.55$. Thus the value of the pion decay constant as
extracted from the bare axial correlation function receives a large
renormalization and we find $F_{\pi}^{R} \approx 130 \,{\rm MeV}$.  We
stress again that the values of $F_\pi$ and $F_{\pi}^{R}$ computed
here are values of the quenched approximation.

The difficulty to extract $\Sigma$ from our data is illustrated in
figure~\ref{axiFig} (left). It shows fits to $t_f = 7$ points
with $F_{\pi}$ as a free parameter, for $\Sigma$ fixed as
$0$ resp.\ $(250~{\rm MeV})^{3}$. Indeed, those two curves
can hardly be distinguished. 

\FIGURE[t]{\epsfig{file=newfit.epsi,angle=270,width=.7\textwidth,clip=}%
\caption{The axial correlation function on a $12^{4}$ lattice at
  $\beta =6$ in the sector $\vnu = 1$ (on the left). The two lines ---
  which are hardly distinguishable in this plot --- are fits of the
  curve predicted by quenched $\chi$PT with the free parameter
  $F_{\pi}$ and $\Sigma = 0 $ resp.\ $(250 \, {\rm MeV})^{3}$. On the
  right we display the values of $F_{\pi}$ obtained from fits over
  $t_{f}$ points around the center at \, $t=6$.\label{axiFig}\label{fig5}}}

Finally, as a consistency check
we also take a look at charge $\vnu =2$. 
Figure~\ref{nu2Fig} shows the data and the predicted curve
for the parameter range that we extracted before from $\vnu =1$.
We see that these curves are compatible
with our data, within the errors.

\FIGURE[t]{\epsfig{file=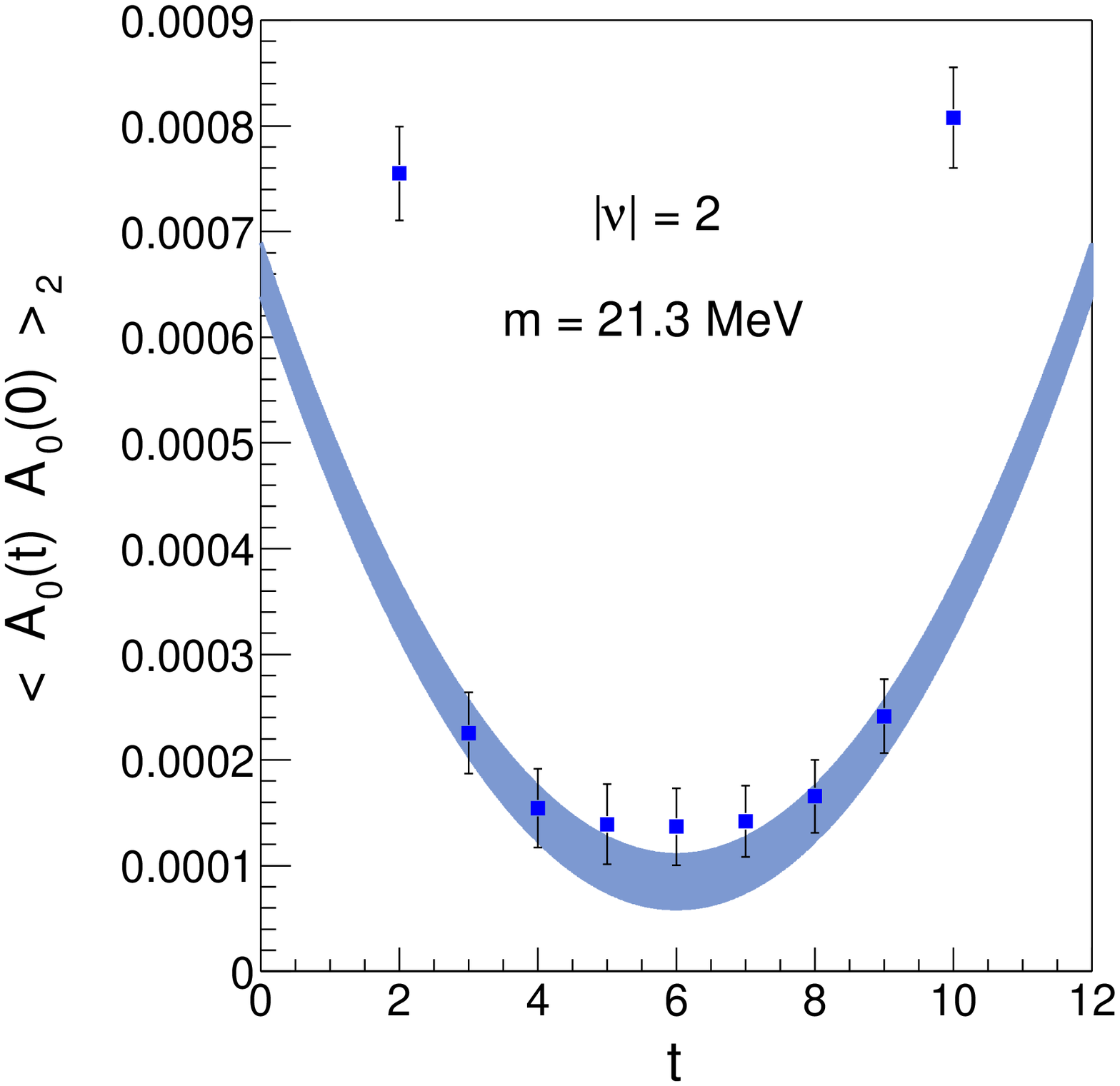,angle=-90,width=.62\textwidth,clip=}%
\caption{Data from a $12^{4}$ lattice at $\beta =6$ in the
sector $\vnu =2$, along with the curves from
eq.~(\ref{AA}). The dark area shows these curves
for the parameters in the range that we determined
from $\vnu =1$. We find agreement with the $\vnu =2$ data
within the errors.\label{nu2Fig}\label{fig6}}}

\section{Conclusions}\label{section4}

We presented a pilot study of the axial correlation function --- as
the simplest example of a meson correlation function --- in the
$\eps$-regime, based on quenched QCD with overlap fermions.  We find
that there are several conditions for running conclusive simulations
with Ginsparg-Wilson fermions in the $\eps$-regime.  The linear size
should obey $L \gsim 1.1~{\rm fm}$, and measurements have to be
performed in a sector of \emph{non-trivial topology.}  The quark mass
should be small for conceptual reasons, but taking it too small causes
technical problems in the measurements.  The above requirement on the
linear size is consistent with an earlier observation based on a
comparison to the Random Matrix predictions for the
eigenvalues~\cite{BJS}.

We observed that the neutral sector, $\nu =0$, is highly unfavorable
for measurements, due to the frequent occurrence of very small
eigenvalues. The corresponding configurations show marked spikes,
which lead to the requirement of a huge statistics of $O(10^{4})$
configurations.

Therefore we worked in the sectors $\vnu =1$ and $2$.  Although our
statistics is still rather modest, we recognize that the data
\emph{can} be fitted to the functions predicted by quenched $\chi$PT.
These fits allow for a stable determination of the quenched value of
$F_{\pi}$.  With our present analysis we were not able to determine
$\Sigma$ since large variations of this parameter hardly changes the
fit to the axial correlation function.  The only exception is the
sector $\nu =0$, see figure~\ref{small-lat}.  However, exactly there
the statistical results are ruined by strong spikes, as we pointed out
before.

As an \emph{outlook}, further information could be obtained by
considering in addition the pseudoscalar-pseudoscalar and the
scalar-scalar correlation function~\cite{TC}.  Then the predictions
involve four free parameters, hence it will be more difficult to
establish a consistent picture.  We also plan to proceed to larger
physical volumes where we expect complete agreement with $\chi$PT.  In
particular, these parameters would then fix directly a one loop
approximation to $\Sigma_{\nu}$ in eq.~(\ref{AA}), which is also a
rough approximation to $\Sigma$.  We also intend to verify the
emerging picture with an alternative Ginsparg-Wilson fermion
formulation~\cite{HF}.

The results of this paper suggest that it is advantageous to work in
fixed topological sectors with relatively large charges, say $\vnu = 2
\dots 5$, in order to suppress the effects of low-lying (non-zero)
eigenvalues.  Hence algorithmic tools or modified gauge actions that
would allow for such simulations are highly desirable and should be
tested in practice to explore the $\epsilon$-regime.

\acknowledgments 

We would like to thank Poul Damgaard, Pilar Hern\'{a}ndez, Heiri
Leutwyler, Martin L\"{u}scher, Rainer Sommer and Sasa Prelovsek for
useful comments.  This work was supported by the DFG through the
SFB/TR9-03 (Aachen-Berlin-Karlsruhe) and in part by the European Union
Improving Human Potential Programme under contracts HPRN-CT-2002-00311
(EURIDICE).  We also thank the John von Neumann-Institute for
Computing and the Konrad-Zuse-Zentrum for providing the necessary
computer resources.

\end{document}